 \documentclass[aps,prl,reprint,preprintnumbers,floatfix,nofootinbib]{revtex4-1}

\usepackage{amsmath}
\usepackage{amssymb}

\usepackage{graphicx}
\usepackage{subcaption}
\usepackage{epstopdf}
\usepackage{dcolumn}
\usepackage{bm}
\usepackage{marvosym}
\usepackage{rotating}

\usepackage{color}
\definecolor{light-gray}{gray}{0.5}
\definecolor{blue}{rgb}{0.0,0.0,1.0}
\definecolor{green}{rgb}{0.0,0.5,0.0}
\definecolor{red}{rgb}{1.0,0.0,0.0}
\definecolor{cyan}{rgb}{0.0,0.75,0.75}
\definecolor{magenta}{rgb}{0.75,0.0,0.75}
\definecolor{yellow}{rgb}{0.75,0.75,0.0}

\newcommand{\avg}[1]{\langle{#1}\rangle}
\newcommand{\sdot}{\cdot}

\newcommand{\grad}{\bm \nabla}
\newcommand{\pd}{\partial}

\begin{document}
\title{The signature of initial conditions on magnetohydrodynamic turbulence}
\author{V. Dallas}
\email{vassilios.dallas@lps.ens.fr}
\author{A. Alexakis}
\affiliation{Laboratoire de Physique Statistique, \'Ecole Normale Sup\'erieure, Universit\'e Pierre et Mari\'e Curie, Universit\'e Paris Diderot, CNRS, 24 rue Lhomond, 75005 Paris, France}

\begin{abstract}
We demonstrate that special correlations in the initial conditions of freely evolving, homogeneous magnetohydrodynamic (MHD) turbulence can lead to the formation of enormous current sheets. These coherent structures are observed at the peak of the energy dissipation rate and are the carriers of long-range correlations despite all the non-linear interactions during the formation of turbulence. Even though the largest scale separation has been achieved at this point, these structures are coherent with a size that spans our computational domain dominating the scaling of the energy spectrum, which follows a $E \propto k^{-2}$ power law. As Reynolds number increases curling of the current sheets, due to Kelvin-Helmhotlz type instabilities and reconnection, modifies the scaling of the energy spectrum from $k^{-2}$ towards $k^{-5/3}$. This transition occurs at the highest Reynolds numbers of direct numerical simulations with resolutions up to $2048^3$ grid points. Finite Reynolds number behaviour is observed due to the initial correlations without reaching a finite asymptote for the energy dissipation rate and with an unexpected scaling between the Taylor and the integral scale Reynolds numbers, i.e. $Re_\lambda \propto Re^{2/3}$. Our results, therefore, demonstrate that even state-of-the-art numerical simulations of the highest resolution can be influenced by the choice of initial conditions and consequently they are inadequate to deduce unequivocally the fate of universality in MHD turbulence.
\end{abstract}

\maketitle

It is generally expected that non-linear interactions in the early stage of the development of turbulence lead to a loss of information associated with the initial conditions (i.e. memory of turbulent fluctuations) \cite{sagautcambon08}. At high enough Reynolds numbers, the maximum value of the energy dissipation rate is expected to asymptote to a finite value and a universal inertial range is expected to be observed in the turbulent energy spectrum with a large population of almost space filling structures of a wide range of length scales \cite{frisch95,biskamp03}. The precise power law scaling of the energy spectrum has implications to the prediction of numerous astrophysical phenomena, such as the heating rates of the solar corona and acceleration of the solar wind \cite{mcintoshetal11}, the transport of mass and energy into the Earth's magnetosphere \cite{sundkvistetal05}, the dynamics of the interstellar medium \cite{gaensleretal11}, etc. However, observations, simulations and theory to date are unable to provide a definitive answer to the power law scaling of the energy spectrum of magnetohydrodynamic (MHD) turbulent flows \cite{k41a,iroshnikov64,kraichnan65,goldreichsridhar95,ngbhattacharjee97,galtieretal00,sauretal02,mullergrappin05,boldyrev06,mininnipouquet07,podestaetal07,beresnyak12}.
Hence, universality in MHD has been questioned by many authors \cite{schekochihinetal08,leeetal10,mininni11,wanetal12,da13a,da13c} in terms of various arguments such as dependence on initial conditions, non-locality, strong anisotropy, lack of self-preservation.

In this letter we demonstrate the impact of memory of initial conditions in freely evolving, homogeneous MHD turbulence. We show that coherent flow structures can actually be the carriers of long-range memory and thus imply long-range effects of initial conditions that can influence the turbulent statistics such as the power law scaling of the energy spectrum. Our results suggest that higher Reynolds number numerical simulations than those feasible at the moment are imperative in order to avoid any finite Reynolds number effects and to be able to infer unequivocally the fate of universality in MHD turbulence.

To demonstrate these points, we perform high resolution simulations of the incompressible MHD equations 
\begin{align}
 (\pd_t - \nu \bm\Delta) \bm u &= (\bm u \times \bm \omega) + (\bm j \times \bm b) - \grad P,
 \label{eq:ns} \\
 (\pd_t - \kappa \bm\Delta) \bm b &=  \grad \times (\bm u \times \bm b), \label{eq:induction}
\end{align}
with $\bm u$ the velocity, $\bm b$ the magnetic field, $\bm \omega \equiv \grad \times \bm u$ the vorticity, $\bm j \equiv \grad \times \bm b$ the current density, $P$ the pressure, $\nu$ the kinematic viscosity and $\kappa$ the magnetic diffusivity. If $\nu = \kappa = 0$, then the total energy $E \equiv \frac{1}{2}\avg{|\bm u|^2 + |\bm b|^2}$, the magnetic helicity $H_b \equiv \avg{\bm u \sdot \bm b}$ and the cross helicity $H_c \equiv \avg{\bm a \sdot \bm b}$ are conserved in time, where $\bm b \equiv \grad \times \bm a$ and $\bm a$ is the solenoidal magnetic potential. Note that the angle brackets $\avg{.}$ here denote spatial averages. Using the pseudo-spectral method, we numerically solve Eqs. \eqref{eq:ns}-\eqref{eq:induction} in a three dimensional periodic box of size $2\pi$, satisfying $\grad \sdot \bm u = \grad \sdot \bm b = 0$. Aliasing errors are removed using the $2/3$ dealiasing rule, i.e. wavenumbers $k_{min}=1$ and $k_{max}=N/3$, where $N$ is the number of grid points in each Cartesian coordinate. For more details on the numerical code see \cite{mpicode05a,hybridcode11}.

The resolutions that we report in this letter range from $N=128$ to $N=2048$ (see Table \ref{tbl:dnsparam}). 
The Reynolds number based on the integral length scale $L \equiv \frac{3\pi}{4}\int k^{-1}E(k)dk/\int E(k)dk$ is $Re = u' L / \nu$ and that based on the Taylor micro-scale $\lambda \equiv (5 \int E(k)dk/\int k^2E(k)dk)^{1/2}$ is $Re_\lambda = u' \lambda / \nu$, where $u' = \avg{|\bm u|^2}^{1/2}$ is the rms velocity. The smallest length scale in our flows is defined based on Kolmogorov scaling $\eta \equiv (\nu^3 / \epsilon)^{1/4}$, where $\epsilon \equiv \nu \avg{|\bm \omega|^2} + \kappa \avg{|\bm j|^2}$ is the total energy dissipation rate. The time we address in our analysis is $t_{peak}$ the moment of maximum $\epsilon$, when the highest scale separation occurs $L \gg \ell \gg \eta$, where $\ell$ is a typical length scale in the inertial range. Thus, the values provided in Table \ref{tbl:dnsparam} correspond to that moment.

Inspired by prior work on MHD Taylor-Green flows \cite{leeetal10,da13a,da13b,da13c}, where $\bm b$ initially satisfied the same Taylor-Green symmetries with $\bm \omega$ and $\bm u$ with $\bm j$, we investigate the effect of initial long-range cross-correlations in MHD turbulent flows. 
So, here we consider initial conditions, which are excited at wave numbers $k = 1$ and $2$ with random phases and we introduce an initial strong cross-correlation between the velocity and the current density by setting $\bm j \propto \bm u$. Initially, our fields are normalised such that the kinetic energy $E_u \equiv \frac{1}{2}\avg{|\bm u|^2}$ and the magnetic energy $E_b \equiv \frac{1}{2}\avg{|\bm b|^2}$ are in equipartition (viz. $E_u = E_b = 0.5$) and all the helicities are zero including the kinetic helicity $H_u \equiv \avg{\bm u \sdot \bm \omega}$. Note that the magnetic Prandtl number is unity (i.e. $\nu = \kappa$) for all the simulations.

\begin{table}[!ht]
  \caption{Numerical parameters and values obtained at the peak of the energy dissipation rate.}
  \label{tbl:dnsparam}
   \begin{ruledtabular}
    \begin{tabular}{*{10}{c}}
     \textbf{N} & $\bm \nu = \bm \kappa$ & $\bm{Re}$ & $\bm{Re_\lambda}$ & $\bm{u'}$ & $\bm{b'}$ & $\bm {k_{max}\eta}$ \\
     \hline
       128 & $5 \times 10^{-3}$ &  178.0 &  68.5 & 0.73 & 1.02 & 1.23 \\
       256 & $3 \times 10^{-3}$ &  273.0 &  94.3 & 0.73 & 1.04 & 1.74 \\
       512 & $1 \times 10^{-3}$ &  639.2 & 166.3 & 0.72 & 0.98 & 1.56 \\
      1024 & $4 \times 10^{-4}$ & 1550.3 & 298.6 & 0.75 & 1.02 & 1.62 \\
      2048 & $2 \times 10^{-4}$ & 2972.5 & 438.0 & 0.73 & 1.04 & 1.98 \\
    \end{tabular}
  \end{ruledtabular}
\end{table}

\begin{figure}[!ht]
  \begin{subfigure}{0.49\textwidth}
   \includegraphics[width=8.5cm]{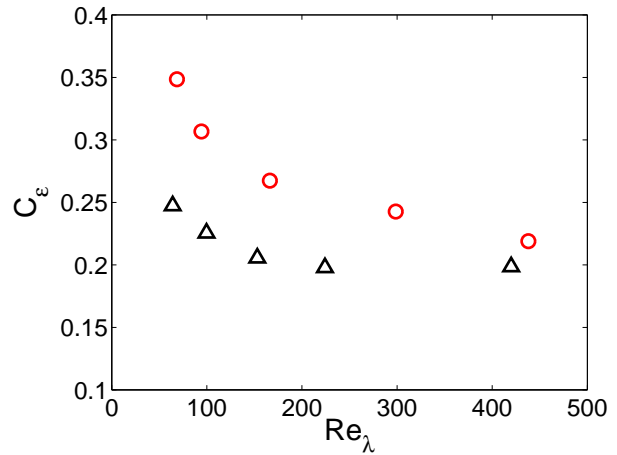}
   \caption{}
  \end{subfigure}
  \begin{subfigure}{0.49\textwidth} 
  \includegraphics[width=8.5cm]{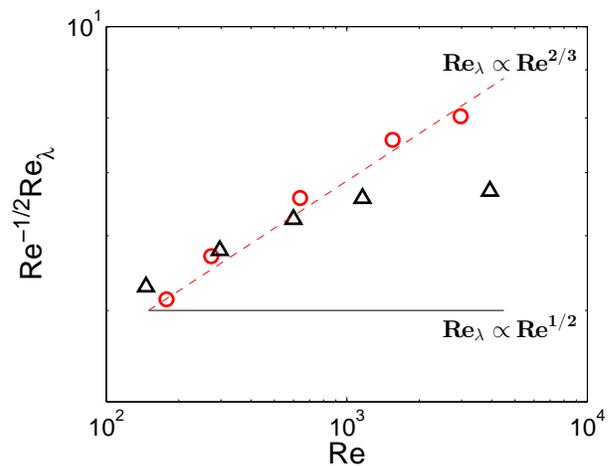}
   \caption{}
  \end{subfigure}
  \caption{(Color online) (a) Dissipation rate coefficient $C_\epsilon$ as a function of $Re_\lambda$. (b) $Re_\lambda$ compensated by $Re^{-1/2}$ versus $Re$. The circles $\bigcirc$ and the triangles $\triangle$ represent our data (see Table \ref{tbl:dnsparam}) and the data from \cite{mininnipouquet09}, respectively.}
  \label{fig:scalings}
 \end{figure}

In this work, we compare our results with the results by Mininni and Pouquet \cite{mininnipouquet09}, who carried out direct numerical simulations (DNS) by superposing Fourier modes with random phases of initially uncorrelated fields for $N=64$ to $N=1536$ (see \cite{mininnipouquet09} for more details). In Fig. \ref{fig:scalings}(a), we plot the dissipation coefficient $C_\epsilon \equiv \epsilon L_0 / u'_0$ for each $Re_\lambda$, where the value of $\epsilon$ is taken at $t_{peak}$ whereas the integral length scale $L_0$ and the rms velocity $u'_0$ are at time zero. The circles represent our runs (see Table \ref{tbl:dnsparam}) and the triangles represent the runs by \cite{mininnipouquet09}.
According to the data by \cite{mininnipouquet09}, $C_\epsilon$ becomes independent of $Re_\lambda$ as it is expected for high enough Reynolds numbers. This is the so-called dissipation anomaly of three-dimensional turbulence \cite{biskamp03}. So, it seems that the beginning of an asymptotic regime has been reached with $C_\epsilon = const$ and $Re_\lambda \propto Re^{1/2}$ for $Re_\lambda > 200$ [see Fig. \ref{fig:scalings}(b)] as expected for a fully developed turbulent flow that obeys Kolmogorov scaling \cite{frisch95}. On the other hand, our runs with initial $\bm j \propto \bm u$ cross-correlation have different high Reynolds number asymptotics without reaching an asymptotic regime for $C_\epsilon$ even for the highest $Re_\lambda$ run with $2048^3$ grid points, which is at the cutting-edge of the current computational capabilities. It is interesting that the $Re_\lambda$ seems to obey a different power law scaling of the form $Re_\lambda \propto Re^{2/3}$ for this range of parameters. So, a question that arises at this point is if this is a manifestation of non-classical scalings or a finite Reynolds number effect.

\begin{figure}[!ht]
  \begin{subfigure}{4.25cm}
   \includegraphics[width=\textwidth]{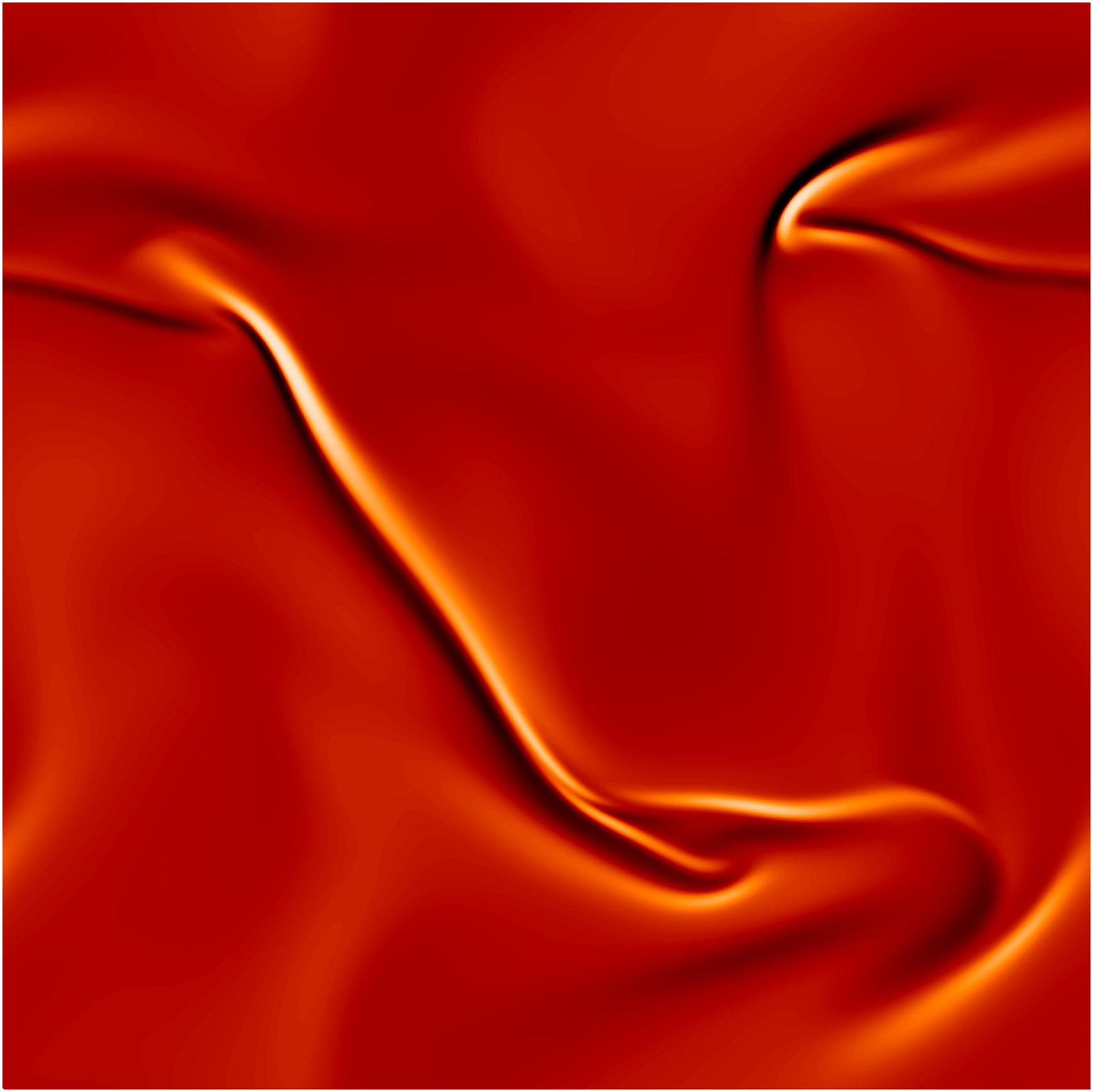}
   \caption{}
  \end{subfigure}
  \begin{subfigure}{4.25cm}
   \includegraphics[width=\textwidth]{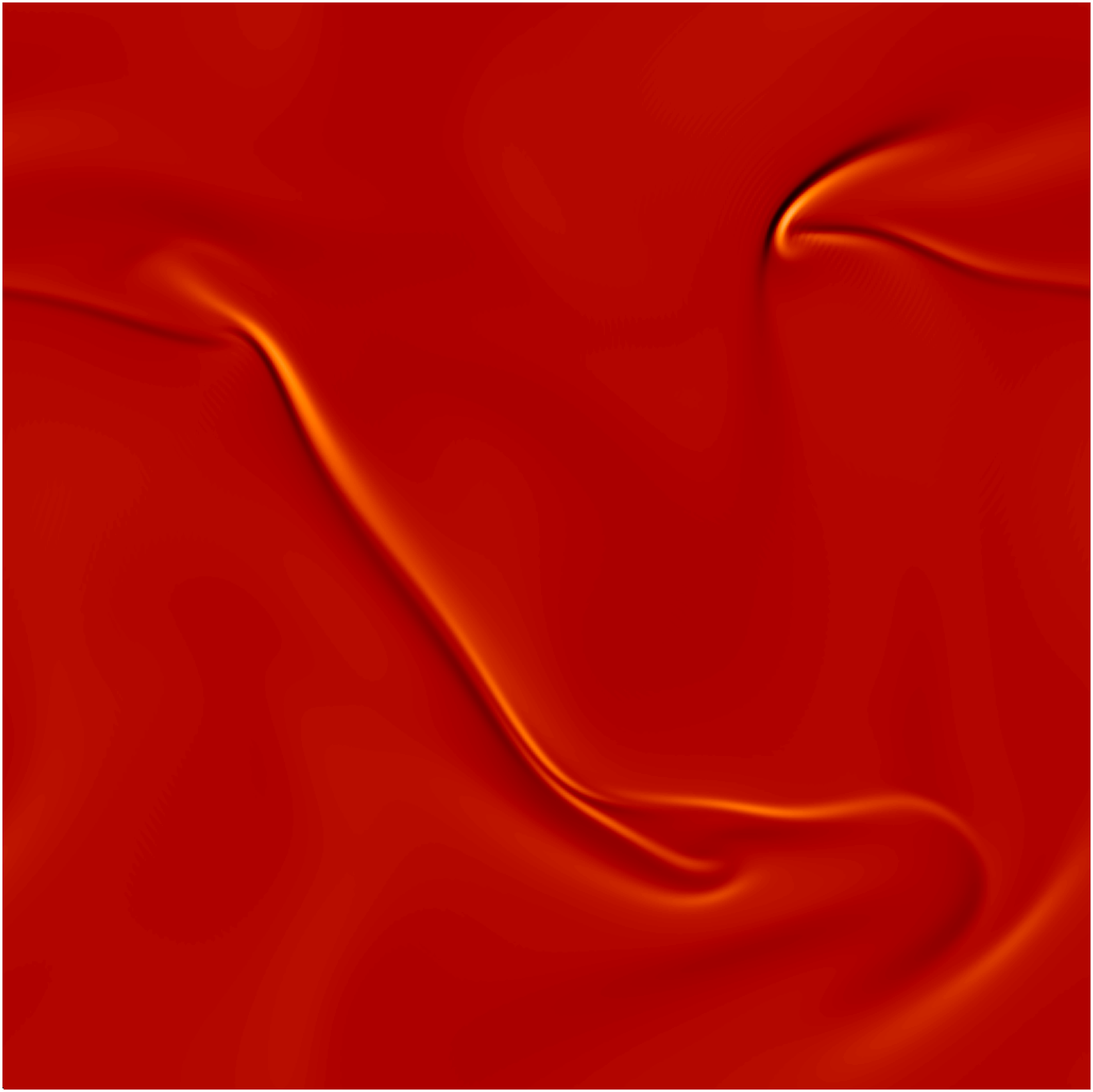}
   \caption{}
  \end{subfigure}
  \begin{subfigure}{4.25cm}
   \includegraphics[width=\textwidth]{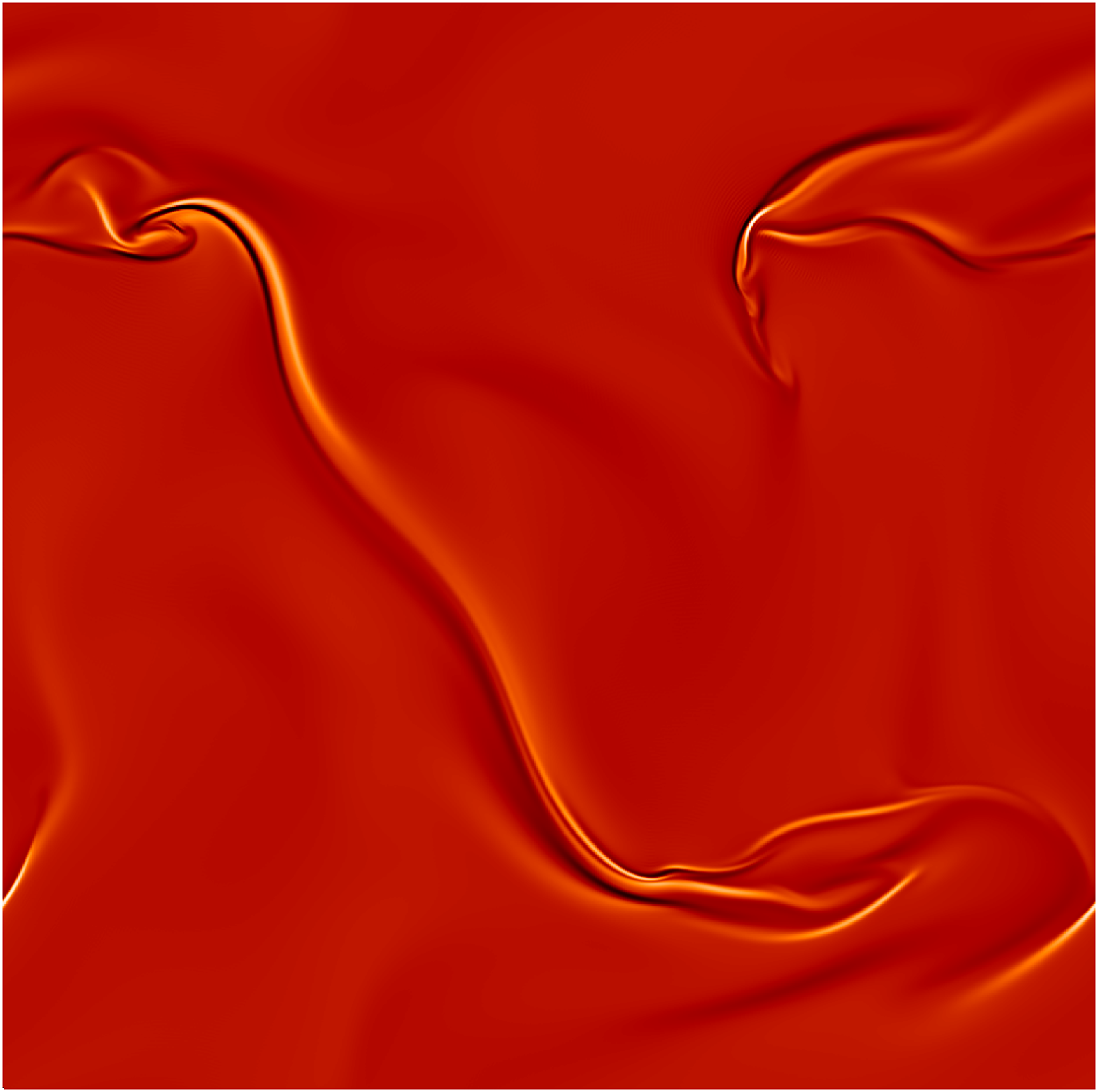}
   \caption{}
  \end{subfigure}
  \begin{subfigure}{4.25cm}
   \includegraphics[width=\textwidth]{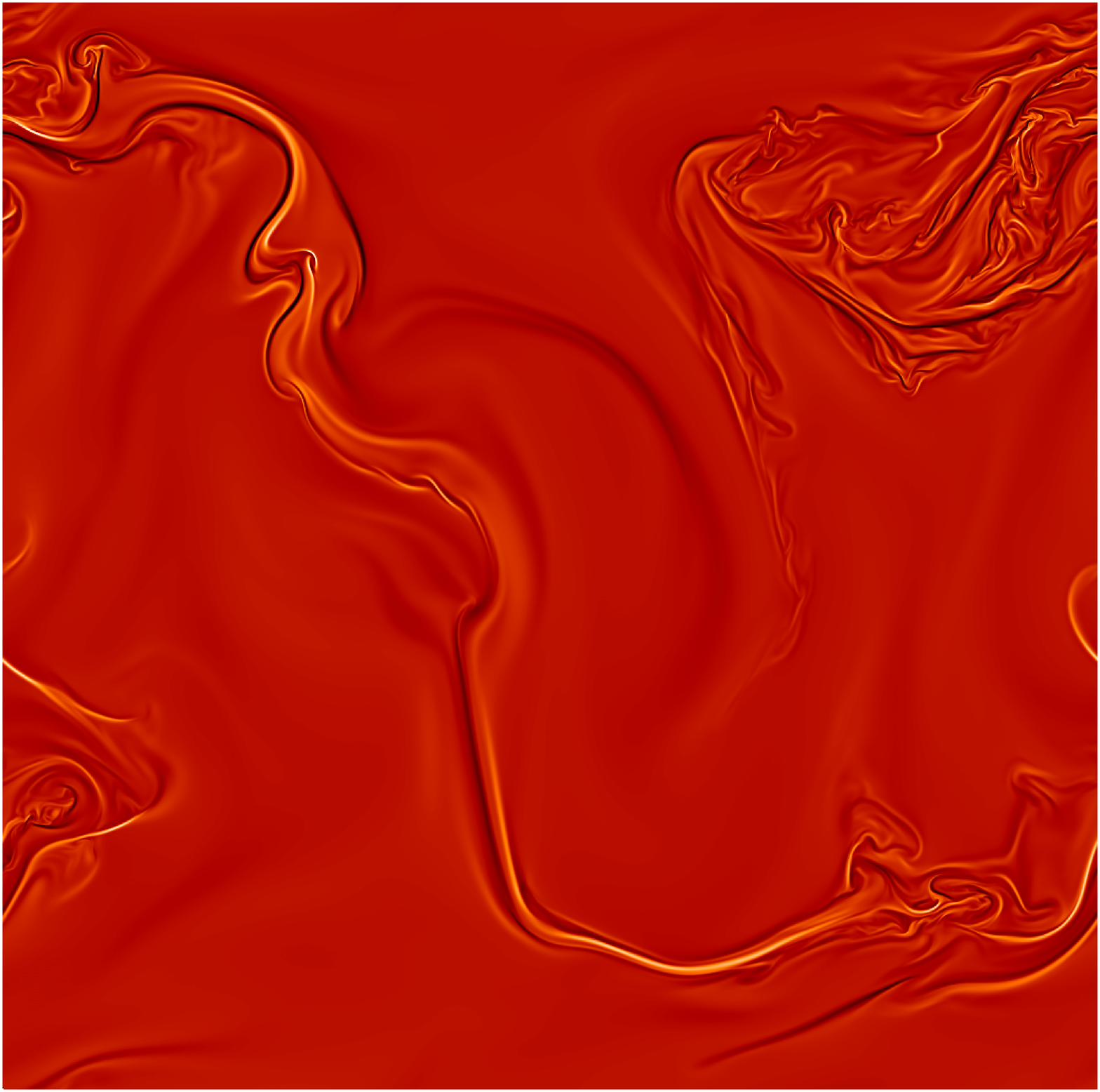}
   \caption{}
  \end{subfigure}
  \caption{(Color online) Current density in a slice of the entire box at the peak of the energy dissipation rate for the runs of Table \ref{tbl:dnsparam}. (a) $Re_\lambda=94.3$, (b) $Re_\lambda=166.3$, (c) $Re_\lambda=298.6$ and (d) $Re_\lambda=438$. High and low intensity regions are denoted by white and black colours, respectively.}
  \label{fig:visual}
 \end{figure}

High resolution DNS of MHD turbulence consist of myriad of intense dissipative sheetlike structures that become more space filling as well as some structures that remain sparse but become thinner as $C_\epsilon$ becomes independent of $\nu$ and $\kappa$ with increasing Reynolds numbers (see also \cite{mininnipouquet09}). Visualisations of the current density amplitude in a slice of the entire computational domain are illustrated in Fig. \ref{fig:visual} at $t_{peak}$ for the four highest $Re_\lambda$ runs of Table \ref{tbl:dnsparam}.
In contrast to conventional DNS, we observe enormous current sheets that span the size of our periodic boxes. Note that these coherent structures are not space filling with thickness of order $\eta$ but spanwise length of order $L$. As $Re_\lambda$ increases, curling of these current sheets takes place as a result of Kelvin-Helmoltz type instabilities until they become unstable and reconnect. Such instabilities have been recently observed in the large population of current sheets of typical random MHD turbulent simulations \cite{mininnipouquet09} but also at the large scales of solar wind observations \cite{hasegawaetal04}. It is remarkable that these sheets are so coherent that the appearance of a large population of smaller space filling intense dissipative structures is observed after reconnection only for the highest $Re_\lambda$ simulation with $2048^3$ grid points.

\begin{figure}[!ht]
  \includegraphics[width=8.5cm]{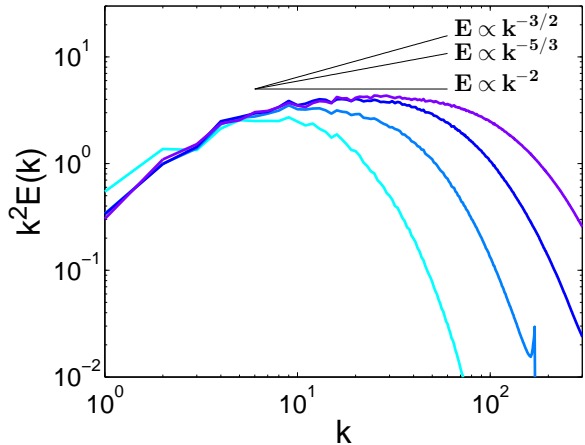}
  \caption{(Color online) The total energy spectrum $E(k)$ compensated by $k^2$ for the four highest $Re_\lambda$ simulations of Table \ref{tbl:dnsparam}.}
  \label{fig:spectra}
 \end{figure}

Figure \ref{fig:spectra} presents the compensated total energy spectra $k^2E(k)$ of the fields shown in Fig. \ref{fig:visual} with the power laws $k^{-2}$, $k^{-5/3}$ and $k^{-3/2}$ also denoted in the plot. These power law scaling exponents are in summary those proposed by the various phenomenologies for MHD turbulence based on weak and strong turbulence arguments both for isotropic and anisotropic energy spectra.
The amplitude of these large scale current sheets is strong enough to dominate the energy spectrum, something that is not typically observed in DNS of freely evolving MHD turbulence with random initial conditions. It is interesting that the total energy spectrum seems to form a $k^{-2}$ scaling at least for the runs that the large scale current sheets are still coherent. As Reynolds number increases the spectrum slowly deviates from the $k^{-2}$ power law towards the $k^{-5/3}$ scaling. Note, however, the absence of a clear power law even for the run with $2048^3$ grid points. At high enough Reynolds numbers, we anticipate that either the $k^{-5/3}$ or the $k^{-3/2}$ will be reached. This result is in agreement with Dallas and Alexakis \cite{da13c}, who demonstrated that the $-2$ power law scaling of the energy spectrum, that originates from magnetic discontinuities corresponding to strong amplitude current sheets in flows with Taylor-Green symmetries \cite{da13b}, will deviate either towards the $-5/3$ or the $-3/2$ scaling exponents at $Re \gg 1$.

\begin{figure}[!ht]
  \includegraphics[width=8.5cm]{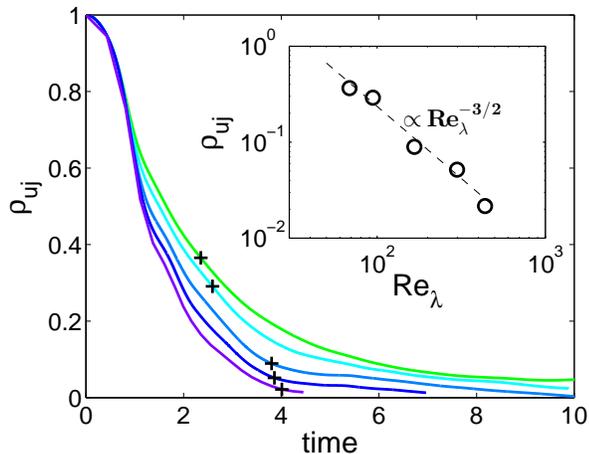}
  \caption{(Color online) Time evolution of the correlation coefficient $\rho_{uj}$ between the velocity and the current density for all the runs of Table \ref{tbl:dnsparam}. The green (light gray) curve denotes the lowest $Re_\lambda$ run and the purple (dark gray) curve the highest $Re_\lambda$ run. The crosses $+$ indicate the value of $\rho_{uj}$ at the time of maximum dissipation rate $t_{peak}$. The inset presents the values of $\rho_{uj}$ at $t_{peak}$ for all the different $Re_\lambda$ runs.}
  \label{fig:rhouj}
 \end{figure}

A measure to quantify the memory of the initial conditions $\bm j \propto \bm u$ is the time evolution of the correlation coefficient between the velocity and the current density $\rho_{uj} = \avg{\bm u \sdot \bm j}/(\avg{|\bm u|^2}\avg{|\bm j|^2})^{1/2}$ that we show in Fig. \ref{fig:rhouj}.
Initially, the velocity and the current density are fully correlated (i.e. $\rho_{uj} = 1$) and then as time evolves $\rho_{uj}$ decays towards zero. Note, however that this decay rate is faster as $Re_\lambda$ increases. The crosses in Fig. \ref{fig:rhouj} indicate the value of $\rho_{uj}$ at the time of the maximum energy dissipation rate. It is clear that for the lower $Re_\lambda$ runs this long-range cross-correlation between $\bm u$ and $\bm j$ is significant with the current sheets having a spanwise length of the order of the box size $2\pi$ and a $k^{-2}$ power law for the energy spectra (see Figs. \ref{fig:visual} and \ref{fig:spectra}). For the highest $Re_\lambda$ runs $\rho_{uj} \rightarrow 0$. Then, we observe the power law of the energy spectrum to start deviating from the $E \propto k^{-2}$ scaling due to the loss of the stability of the current sheets, which consequently break down into a large population of smaller scale structures. Furthermore, the inset of Fig. \ref{fig:rhouj} presents the values of $\rho_{uj}$ at $t_{peak}$ for the different $Re_\lambda$ runs of Table \ref{tbl:dnsparam}. This data exhibits a power law of the form $\rho_{uj} \propto Re_\lambda^{-3/2}$ demonstrating the rate which MHD turbulence loses memory of initial conditions.

In summary, we performed high resolution simulations of freely evolving MHD turbulence with initial cross-correlation of the type $\bm j \propto \bm u$. We focus at the time when the peak of energy dissipation rate is reached, which is when the largest scale separation occurs. At this point, we observe current sheets with a spanwise length of the order of the box size $2\pi$ and thickness of the order of the Kolmogorov scale $\eta$. These structures remain coherent in the flow even for our highest $Re_\lambda$ runs. Their amplitude is strong enough to dominate the energy spectrum particularly for our lower resolution runs with a $E \propto k ^{-2}$ power law scaling as the best fit. As $Re_\lambda$ increases, Kelvin-Helmoltz type instabilities entail the curling of the enormous current sheets that spontaneously break down to smaller scales only for our run with $2048^3$ grid points, modifying the scaling of the energy spectrum from $k ^{-2}$ towards $k ^{-5/3}$. Therefore, we deduce that the $k^{-2}$ energy spectrum manifests as a finite Reynolds number effect and hence the debate on universality in MHD turbulence returns back to the distinction between the $k^{-5/3}$ and the $k^{-3/2}$ scalings at high enough Reynolds numbers.

Overall, our work emphasizes the importance of memory of initial conditions in freely evolving MHD turbulence. The rate which the cross-correlation weakens is relatively slow (i.e. $\rho_{uj} \propto Re_\lambda^{-3/2}$) for the resolutions that can be achievable using todays most powerful supercomputers. We anticipate that for higher $Re_\lambda$ the energy dissipation rate will reach an asymptote and the scaling $Re_\lambda \propto Re^{2/3}$ will be modified.

It is known that the amount of cross-correlation between the velocity and the magnetic fields is flow dependent in astrophysical phenomena. So, observations should investigate if long-range cross-correlations of the type we considered in this study can influence the turbulent statistics. For example, 
such cross-correlations might be important in the magnetosphere of Jupiter, where indications of $k^{-2}$ scaling of the magnetic energy spectrum are reported \cite{sauretal02}.

Eventually, it is apparent that there is an urgent need for higher resolutions in numerical simulations of MHD turbulence to be able to deduce undoubtedly the fate of universality in the high Reynolds number limit. The fact that MHD turbulence remembers certain information from the initial conditions despite all of the complex non-linear interactions is an important issue that needs further investigation.
\begin{acknowledgements}
We acknowledge P. D. Mininni and A. Pouquet for their data. V.D. acknowledges the financial support from EU-funded Marie Curie Actions--Intra-European Fellowships (FP7-PEOPLE-2011-IEF, MHDTURB, Project No. 299973). The computations were performed using the HPC resources from GENCI-TGCC-CURIE (Project No. x2013056421) and PRACE-FZJ-JUQUEEN (Project name PRA068).
\end{acknowledgements}

\bibliography{references}
\end{document}